\documentclass[twocolumn]{aastex6}

\begin{document}
 \title{Imaging the 44 AU Kuiper Belt-analogue debris ring around HD 141569A with GPI polarimetry} 
\author{Juan Sebasti\'an Bruzzone\altaffilmark{1}, Stanimir Metchev\altaffilmark{1,2,3}, Gaspard Duch\^ene\altaffilmark{4,5}, Maxwell A. Millar-Blanchaer\altaffilmark{6}, Ruobing Dong\altaffilmark{7}, Jason J. Wang\altaffilmark{4}, James R. Graham\altaffilmark{4}, Johan Mazoyer\altaffilmark{8}, Schuyler Wolff\altaffilmark{8}, S. Mark Ammons\altaffilmark{9}, Adam C. Schneider\altaffilmark{7}, Alexandra Z. Greenbaum\altaffilmark{10}, Brenda C. Matthews\altaffilmark{11,12}, Pauline Arriaga\altaffilmark{13}, Vanessa P. Bailey\altaffilmark{6}, Travis Barman\altaffilmark{14}, Joanna Bulger\altaffilmark{15}, Jeffrey Chilcote\altaffilmark{16,17}, Tara Cotten\altaffilmark{18}, Robert J. De Rosa\altaffilmark{4}, Rene Doyon\altaffilmark{19}, Michael P. Fitzgerald\altaffilmark{13}, Katherine B. Follette\altaffilmark{20}, Benjamin L. Gerard\altaffilmark{11,12}, Stephen J. Goodsell\altaffilmark{21}, Pascale Hibon\altaffilmark{22}, Justin Hom\altaffilmark{7}, Li-Wei Hung\altaffilmark{13}, Patrick Ingraham\altaffilmark{23}, Paul Kalas\altaffilmark{4,24}, Quinn Konopacky\altaffilmark{25}, James E. Larkin\altaffilmark{13}, Bruce Macintosh\altaffilmark{16}, J\'er\^ome Maire\altaffilmark{25}, Franck Marchis\altaffilmark{24}, Christian Marois\altaffilmark{11,12}, Katie M. Morzinski\altaffilmark{26}, Eric L. Nielsen\altaffilmark{24,16}, Rebecca Oppenheimer\altaffilmark{27}, David Palmer\altaffilmark{9}, Rahul Patel\altaffilmark{28}, Jennifer Patience\altaffilmark{7}, Marshall Perrin\altaffilmark{8}, Lisa Poyneer\altaffilmark{9}, Laurent Pueyo\altaffilmark{8}, Abhijith Rajan\altaffilmark{7}, Julien Rameau\altaffilmark{19}, Fredrik T. Rantakyr\"o\altaffilmark{29}, Dmitry Savransky\altaffilmark{30}, Anand Sivaramakrishnan\altaffilmark{8}, Inseok Song\altaffilmark{18}, Remi Soummer\altaffilmark{8}, Sandrine Thomas\altaffilmark{23}, J. Kent Wallace\altaffilmark{6}, Kimberly Ward-Duong\altaffilmark{20}, Sloane Wiktorowicz\altaffilmark{31}}
\affil{$^1$Department of Physics and Astronomy, The University of Western Ontario, London, ON N6A 3K7, Canada \\ 
 $^2$Institute of Earth and Space Exploration, London, ON N6A 3K7, Canada\\
 $^{3}$Department of Physics \& Astronomy, Stony Brook University, Stony Brook, NY, 11794--3800,USA\\
 $^4$Astronomy Department, University of California, Berkeley, CA 94720, USA \\
 $^5$Universit\'e Grenoble Alpes, CNRS, IPAG, Grenoble, F-38000, France\\
 $^6$Jet Propulsion Laboratory, California Institute of Technology Pasadena CA 91109, USA\\
 $^{7}$ School of Earth and Space Exploration, Arizona State University, PO Box 871404, Tempe, AZ 85287, USA \\
 $^{8}$ Space Telescope Science Institute, Baltimore, MD 21218, USA \\
 $^{9}$ Lawrence Livermore National Laboratory, Livermore, CA 94551, USA \\
 $^{10}$ Department of Astronomy, University of Michigan, Ann Arbor, MI 48109, USA \\
 $^{11}$ National Research Council of Canada Herzberg, 5071 West Saanich Rd, Victoria, BC, V9E 2E7, Canada \\
 $^{12}$ University of Victoria, 3800 Finnerty Rd, Victoria, BC, V8P 5C2, Canada \\
 $^{13}$ Department of Physics \& Astronomy, University of California, Los Angeles, CA 90095, USA \\
 $^{14}$ Lunar and Planetary Laboratory, University of Arizona, Tucson AZ 85721, USA \\
 $^{15}$ Subaru Telescope, NAOJ, 650 North A{'o}hoku Place, Hilo, HI 96720, USA \\
 $^{16}$ Kavli Institute for Particle Astrophysics and Cosmology, Stanford University, Stanford, CA 94305, USA \\
 $^{17}$ Department of Physics, University of Notre Dame, 225 Nieuwland Science Hall, Notre Dame, IN, 46556, USA\\
 $^{18}$ Department of Physics and Astronomy, University of Georgia, Athens, GA 30602, USA \\
 $^{19}$ Institut de Recherche sur les Exoplan{\`e}tes, D{\'e}partement de Physique, Universit{\'e} de Montr{\'e}al, Montr{\'e}al QC, H3C 3J7, Canada \\
 $^{20}$ Physics and Astronomy Department, Amherst College, 21 Merrill Science Drive, Amherst, MA 01002, USA\\
 $^{21}$ Gemini Observatory, 670 N. A'ohoku Place, Hilo, HI 96720, USA \\
 $^{22}$ European Southern Observatory, 3107 Alonso de Córdova, Vitacura, Santiago, Chile\\
 $^{23}$ Large Synoptic Survey Telescope, 950N Cherry Ave., Tucson, AZ 85719, USA \\
 $^{24}$ SETI Institute, Carl Sagan Center, 189 Bernardo Ave., Mountain View CA 94043, USA \\
 $^{25}$ Center for Astrophysics and Space Science, University of California San Diego, La Jolla, CA 92093, USA \\
 $^{26}$ Steward Observatory, University of Arizona, Tucson, AZ 85721, USA \\
 $^{27}$ Department of Astrophysics, American Museum of Natural History, New York, NY 10024, USA \\
 $^{28}$Infrared Processing and Analysis Center, California Institute of Technology, Pasadena, CA 91125, USA \\
 $^{29}$ Gemini Observatory, Casilla 603, La Serena, Chile \\
 $^{30}$ Sibley School of Mechanical and Aerospace Engineering, Cornell University, Ithaca, NY 14853, USA \\
 $^{31}$ Department of Astronomy, UC Santa Cruz, 1156 High St., Santa Cruz, CA 95064, USA }

\email{smetchev@uwo.ca} 
\begin{abstract}
We present the first polarimetric detection of the inner disk component around the pre-main sequence B9.5 star HD 141569A. Gemini Planet Imager $H$-band (1.65$\mu$m) polarimetric differential imaging reveals the highest signal-to-noise ratio detection of this ring yet attained and traces structure inwards to $0\farcs25$ (28 AU at a distance of 111 pc). The radial polarized intensity image shows the east side of the disk, peaking in intensity at $0\farcs40$ (44 AU) and extending out to $0\farcs9$ (100 AU). There is a spiral arm-like enhancement to the south, reminiscent of the known spiral structures on the outer rings of the disk. The location of the spiral arm is coincident with $^{12}$CO J=3\textendash2 emission detected by ALMA, and hints at a dynamically active inner circumstellar region. Our observations also show a portion of the middle dusty ring at $\sim$220 AU known from previous observations of this system. We fit the polarized $H$-band emission with a continuum radiative transfer Mie model. Our best-fit model favors an optically thin disk with a minimum dust grain size close to the blow-out size for this system: evidence of on-going dust production in the inner reaches of the disk.  The thermal emission from this model accounts for virtually all of the far-infrared and millimeter flux from the entire HD~141569A disk, in agreement with the lack of ALMA continuum and CO emission beyond $\sim$100~AU.  A remaining 8--30$\micron$ thermal excess a factor of $\sim$2 above our model argues for a yet-unresolved warm innermost 5--15 AU component of the disk.
\end{abstract}

\section{Introduction}
Debris disks are established laboratories to study planet formation and evolution. Planetesimals and infant planets interact with the dusty disk, and create gaps, asymmetries, offsets and local enhancements through various dynamical mechanisms that help infer their presence (e.g., $\beta$ Pictoris disk and planet, \citealt{Lagage94},  \citealt{Kalas95}, \citealt{Lagrange10}). Of special interest are those young disks (up to 40 Myr) transitioning between the  protoplanetary disk and debris disk stage. These ``hybrid disks'' contain gas that can be either primordial or secondary \citep[e.g.,][]{moor_etal17}, and often show a significant deficit in near-IR or mid-IR emission, or an inner clearing in resolved images (e.g., HD~163296, \citealt{doucet_etal06}; 49 Ceti, \citealt{Hughes2008}; HD 21997, \citealt{Kospal13}). The study of hybrid disks is attractive because of the implications for gas giant planet formation and gas-dust interaction models in nascent planetary systems. 

The star HD 141569A ($V=7.12$ mag, \citealt{Tycho2}; $H=6.86$ mag, \citealt{Cutri03}) is a well-known hybrid disk host. HD 141569A is a young $5\pm3$ Myr \citep{Weinberger00} B9.5V star \citep{Jascheck92} at a distance of $110.5 \pm 0.5$ pc \citep{GAIADR2} with $L_{\text{IR}}/L_{\text{star}}\sim8\times 10^{-3}$ \citep{Sylvester96} and two low mass stellar companions, $B$ and $C$, at roughly 9\arcsec\  \citep{Weinberger00}. Early 12\textendash100 $\mu$m photometry with IRAS indicated a population of $\sim 100$ K circumstellar dust  at an estimated distance of 47\textendash 60 AU \citep{Jaschek86, Walker88}.

The first high-contrast coronographic images in scattered light with HST/NICMOS at $\sim 1.6\mu$m (F160W) revealed a bright dusty disk inclined at $53^{\circ}\pm5^{\circ}$ and position angle (PA) of $355^{\circ} \pm 1^{\circ}$ \citep{Augereau99, Weinberger99}. \cite{Weinberger99} reported a disk extending out to 4\arcsec\ with two conspicuous nested rings peaking at 2\farcs0 (220 AU) and 3\farcs3 (360 AU; along the semi-major axis of the disk), separated by a $1\arcsec$-wide gap devoid of scattering material. In this work, we refer to the 220 AU and 360 AU rings as the middle and outer rings, respectively. The brighter eastern side of the system is likely nearer to us under the assumption of preferentially forward scattering by dust \citep{Weinberger99}. Subsequently, HST optical coronagraphic observations with STIS \citep[365--806 \micron, 50CCD;][]{Mouillet01} and with ACS at 430 nm (F435W), 590 nm (F606W) and 830 nm  \citep[F814W;][]{Clampin03} revealed asymmetries, and two prominent tightly wound spiral substructures: an inner arm between $1\farcs8$\textendash$2\farcs2$ (200 AU\textendash240 AU) and an outer broad ring at 3$\farcs0$\textendash$4\farcs$0 (330 AU\textendash440 AU). In a new analysis of archival Gemini/NICI and HST/NICMOS data, \cite{Mazoyer16} report a split in two fine rings in the eastern part of the disk and show that the $2\arcsec$ ring shows a small 0\farcs03 offset relative to the central star.
 
\cite{Marsh02} were the first to  suggest a peak in the dust optical depth inwards of $\sim 70$ AU from mid-IR imaging with Keck/MIRLIN \citep{Ressler94}. High-contrast coronagraphic observations in the near-IR with VLT/SPHERE further revealed the presence of a third inner ring at $\sim44$ AU \citep{Perrot16}. This resolved inner disk component is also seen as an arc-like rim by \cite{Konishi16} in optical broadband light with HST/STIS, and marginally detected by \cite{Currie16} with Keck/NIRC2 at \emph{L$_{\text{p}}$}. \cite{Mawet17} confirm the detection of the inner disk component around HD 141569A in \emph{L$_{\text{p}}$}-band scattered light with a vortex coronagraph in Keck/NIRC2.  
The combination of optical/near-IR scattered light and 870$\mu$m/2.9 mm ALMA observations \citep{White16, White18} limit the outer radius of the inner disk to $\sim$55~AU.

HD141559A is also known to be a gas-rich disk on large scales, with a total estimated mass in the 13--200 $M_{\oplus}$ (0.39--$6.0\times10^{-4}M_\odot$) range \citep{Zuckerman95, Jonkheid06, Thi14, Flaherty16}.
CO ro-vibrational emission lines trace the existence of the gas between 17--500~AU 
\citep{Brittain03, Goto06, Flaherty16, White16}. However, spatially-resolved ALMA $^{12}$CO ($J = 3-2$) observations reveal that the $<$210~AU region of the disk may contain only a fraction, $\sim$1.5$M_{\oplus}$ ($4.5\times10^{-6}M_\odot$), of this gas mass \citep{White16}, and that the inner $<$50~AU hold only tenuous amounts of CO gas.
 
We present the first $H$-band polarimetric observations of the inner disk of HD 141569A (Section \ref{Sobs}). We use polarimetric differential imaging (PDI) with the Gemini Planet Imager \citep[GPI,][]{Macintosh14}. PDI excels in high-contrast sensitivity to polarized circumstellar emission, as it eliminates the need of further PSF subtraction that can hamper the detection of extended emission. Our PDI observations resolve the inner disk into a ring with polarized intensity peaking at 44 AU and extending inwards to 0$\farcs$25 (28 AU; Section \ref{results}). We model the linear polarized intensity image to derive the physical parameters of the disk (Section \ref{models}). We also compare the predicted thermal emission from our best-fit model against the SED assembled from previous studies (Section \ref{discussion}). We present our conclusions in Section \ref{conclusions}.     


\section{Observations and Data Reduction}\label{Sobs}
We observed HD 141569A on 2014 March 22 UT in polarimetry mode at $H$ band during GPI commissioning at the Gemini South Telescope. We acquired 50 frames of 60 seconds each over $45^{\circ}$ of parallactic field rotation at an average airmass of 1.12. Between each observation the half wave plate (HWP) modulator was rotated in 22.5$^{\circ}$ steps. The HWP introduces modulation in the signal and thereby allows for reconstruction of the Stokes vector later in the reduction steps. During these observations, the average Gemini Differential Image Motion Monitor (DIMM) seeing at Cerro Pach\'on was 0\farcs70. 

We reduced the data with the GPI Data Reduction Pipeline \citep[GPI DRP; ][]{Marie14, Perrin14} following the procedure described in \cite{Perrin15} and \cite{Max15}. This starts with dark subtraction, correction for instrument flexure, microphonics noise, and bad pixels. Each frame is then assembled into a polarization datacube, where the third dimension comprises two image slices, each containing one of the two orthogonal polarization states yielded by the Wollaston prism analyzer. 
Each datacube is divided by a Gemini Facility Calibration Unit (GCAL) flat field, for throughput correction across the field. We correct for instrumental polarization as outlined in \cite{Max15}.

The position of the central star is determined using GPI's four fiducial satellite spots \citep{Wang14}. In broad-band images the satellite spots are smeared radially outwards from the central star and form bright elongated streaks that can be used to estimate the location of the obscured star \citep{Pueyo15}. Following \cite{Wang14}, we use a technique that implements a Radon transform of the flux distribution in each polarized image to compute the line integral over all lines passing through an initial guess for the position of the star. We sum the squares of all line integrals, and repeat the procedure for the next guess for the stellar position within a small search box. The point within the search box that contains the most light pointing towards it corresponds to the stellar position.  This way, we attain the position of the obscured central star to $\sim$1~mas precision \citep{Wang14}.  We then perform a double differencing between the two polarization states to correct for non-common path errors \citep{Perrin15}.

We use the series of differenced datacubes obtained at different HWP angles to construct a single Stokes $[I,Q,U,V]$ datacube, the 2D slices of which hold the total intensity, linear and circular polarization information for the entire observation sequence.  The details of this procedure are presented in Appendix B.2 of \citet{Perrin15}.
Because GPI's HWP is not exactly one half wave at all wavelengths, GPI is only weakly sensitive to circular polarization, Stokes $V$. Thus we disregard the Stokes $V$ cube slice. Afterwards, following \cite{Schmid06}, the Stokes cube was transformed into the radial Stokes convention $[I,Q_{r},U_{r},V]$. In this formalism, positive values of Stokes $Q_{r}$ represent linear polarization perpendicular to the radial direction from the star, while negative values represent polarization parallel to the radial direction. For Rayleigh-like scatterers in an optically thin debris disk, no flux is expected in the Stokes $U_{r}$ image as only the perpendicular macroscopic polarization state (azimuthal polarization) prevails \citep{Max15}. Thus, light from single scattering events by dust grains will lead to positive values in Stokes $Q_{r}$. However, we note that multiple scattering in optically thick disks could have a Stokes $U_{r}$ signal of a few percent of the Stokes $Q_{r}$ signal \citep{Canovas15}, which is below the sensitivity of our observations.

Our final reduction step it to flux-calibrate the data. Following the procedure outlined in \cite{Hung16}, and adopting $1.85 \pm 0.07$ Jy as the $H-$band flux of HD 141569A from 2MASS \citep{Cutri03}, we obtain a conversion factor of $\left(1.05\pm 0.06\right)\times 10^{-8}$ Jy ADU$^{-1}$ coadd$^{-1}$. The final $Q_r$ and $U_r$ images are shown in Figure~\ref{obsGPI}, and an SNR map of the $Q_r$ image is shown in Figure~\ref{snrmap}.

In addition to the Stokes $Q_{r}$ image, we also reduced the total intensity image (Stokes $I$ slice) for the entire sequence with \texttt{pyKLIP} \citep{Wang15KLIP} which implements the Karhunen-Lo\`eve Image Projection algorithm \citep[KLIP,][]{Soummer12} for optimal PSF subtraction.  That reduced image is shown in Figure~\ref{JasonKLIP}.

Immediately after the polarimetry sequence was completed we also acquired integral field spectroscopic (IFS; non-polarimetric) observations of HD 141569A with GPI in the $H$ band. The IFS data contain an independent measurement of the total intensity flux from the disk. The sequence comprised 32 frames of 60 seconds each. The sequence started at an airmass of 1.13, with an average Gemini DIMM seeing of 1\farcs01 and $27^{\circ}$ of cumulative field rotation. We reduced the data and assembled the spectral datacubes with the standard recipes provided in the GPI DRP. The entire reduced dataset was then processed with \texttt{pyKLIP}. However, unlike our polarimetric observations, this shorter IFS sequence did not reveal the disk, and is not shown here.

\section{Results}\label{results}
We present the imaging results in polarized and total intensity in Section \ref{respol} and \ref{stotintens}, respectively. We revisit these further in Sections \ref{discussion1}-\ref{discussionspiral} in the context of previous observations and a scattered-light model of the disk (Section \ref{models}).
\subsection{Polarized Intensity H-band Image}\label{respol}
\begin{figure*}[ht!]
 \begin{center}
  \includegraphics[width=7.1in]{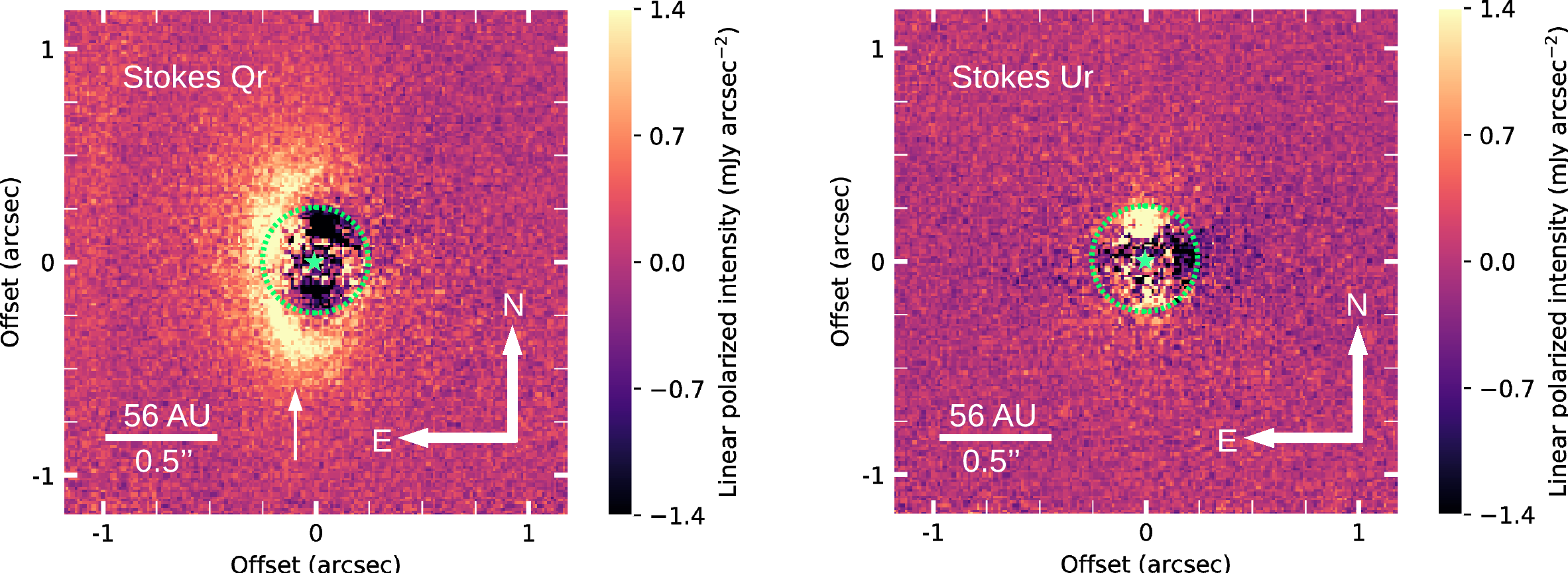}
 \end{center}
\caption{Observations of the HD 141569A dusty disk in $H$-band linear polarization with GPI in a total of 50 minutes of integration. {\it Left:} Linear polarization intensity $Q_{r}$ image. {\it Right:} Stokes $U_{r}$ image shown on the same color scale. An arrow points to the location of the surface brightness enhancement -- an arc feature (see Figure \ref{fig_mirrored_subtractions}) -- to the south. A star symbol marks the position of the star, and a circular aperture of radius $0\farcs25$ centered on the star indicates the region affected by uncorrected instrumental polarization. Beyond this region, the $U_{r}$ image scatters uniformly around zero flux, which indicates that the dust seen in the Stokes $Q_{r}$ is optically thin. The previously known middle ring is visible at low surface brightness at 1\arcsec $~$to the east in the Stokes $Q_{r}$ image.}\label{obsGPI}
\end{figure*}

The $H$-band Stokes $Q_{r}$ image in Figure \ref{obsGPI} shows the first polarized light detection of the HD 141569A inner disk. We clearly resolve the eastern portion of the disk: likely the result of predominantly forward scattering by micron-size dust grains. This is dictated by the combined effects of phase function and polarization dependence with scattering angle. Most known dust compositions preferentially scatter light in the forward direction for dust particles a few times larger than the wavelength of the scattered light \citep{Hulst1957}. For such particles, optically-thin Mie models also suggest that polarized intensity curves peak at scattering angles $\leq 90^{\circ}$ \citep{Perrin15}. 

The emission peaks to the east at a semi-major axis of $0\farcs40$ (44 AU) and extends out to $\sim 0\farcs9$ (100 AU), where it blends into the background. At $1\arcsec$ to the east the Stokes $Q_{r}$ surface brightness increases again revealing the previously detected middle dusty ring at $220$ AU. The clearing between the inner disk and the middle ring indicates a region deficient in scattering material, assuming that the disk is optically thin. 

The western part of the Stokes $Q_r$ image shows no significant emission.  We set a one-sigma lower limit of 4.0 on the ratio of forward to backward scattering in polarized intensity.  If the polarization phase function is symmetric around 90$\degr$, then this also sets a lower limit on the forward to backward scattering intensity.  However, we do not anticipate this to  generally be true, as evidenced by the unusual case of the HR~4796A debris disk \citep{Perrin15, Milli17}, and by theoretical projections for polarization phase functions in \citet{Canovas15}.

\begin{figure}[ht!]
 \centering
  \includegraphics[width=3.2in]{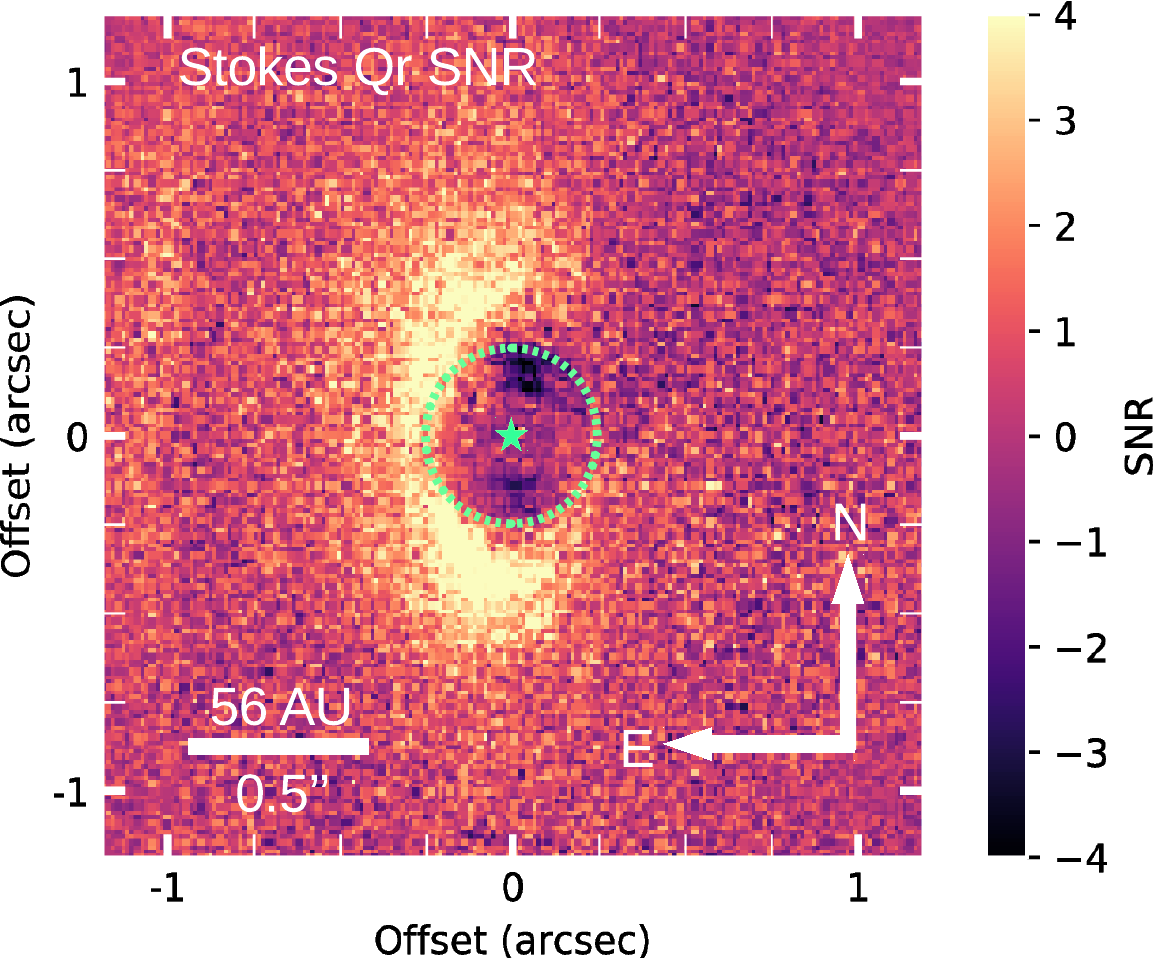}
  \caption{$H$-band Stokes $Q_{r}$ SNR map of HD 141569A. Each point on the map is constructed by dividing the $Q_{r}$ image by the standard deviation of all pixels within a 3 pixel-wide annulus in the $U_{r}$ image at the same angular separation from the star.}\label{snrmap}
\end{figure}

In Section \ref{discussionspiral} we further discuss a residual arc-like structure to the south, which is also detectable as a brightness enhancement in the Stokes $Q_r$ image (Figure \ref{obsGPI}, left panel). Uncorrected instrumental polarization affects the signal within $0\farcs25$ of the star.  We have delimited this region by a dashed circle in Figures~\ref{obsGPI}--\ref{JasonKLIP}, and have excluded it from our analysis. 

No coherent structures are observed in the Stokes $U_{r}$ image to indicate significant optical depth. Hence, the Stokes $U_{r}$ image can be used as a noise map for our measurements and it reassures the astrophysical nature of the Stokes $Q_{r}$ emission: specifically the polarized morphology exterior to $0\farcs25$ and the middle ring near the edge to the east. Figure \ref{snrmap} shows an SNR map created by dividing the $Q_{r}$ image at every position by local noise estimated as the standard deviation of pixels within a 3 pixel-wide annulus in the Stokes $U_{r}$ image at the same angular separation from the star. 

\subsection{Total Intensity H-band Image}\label{stotintens}

\begin{figure}[ht!]
 \center
  \includegraphics[width=3.2in]{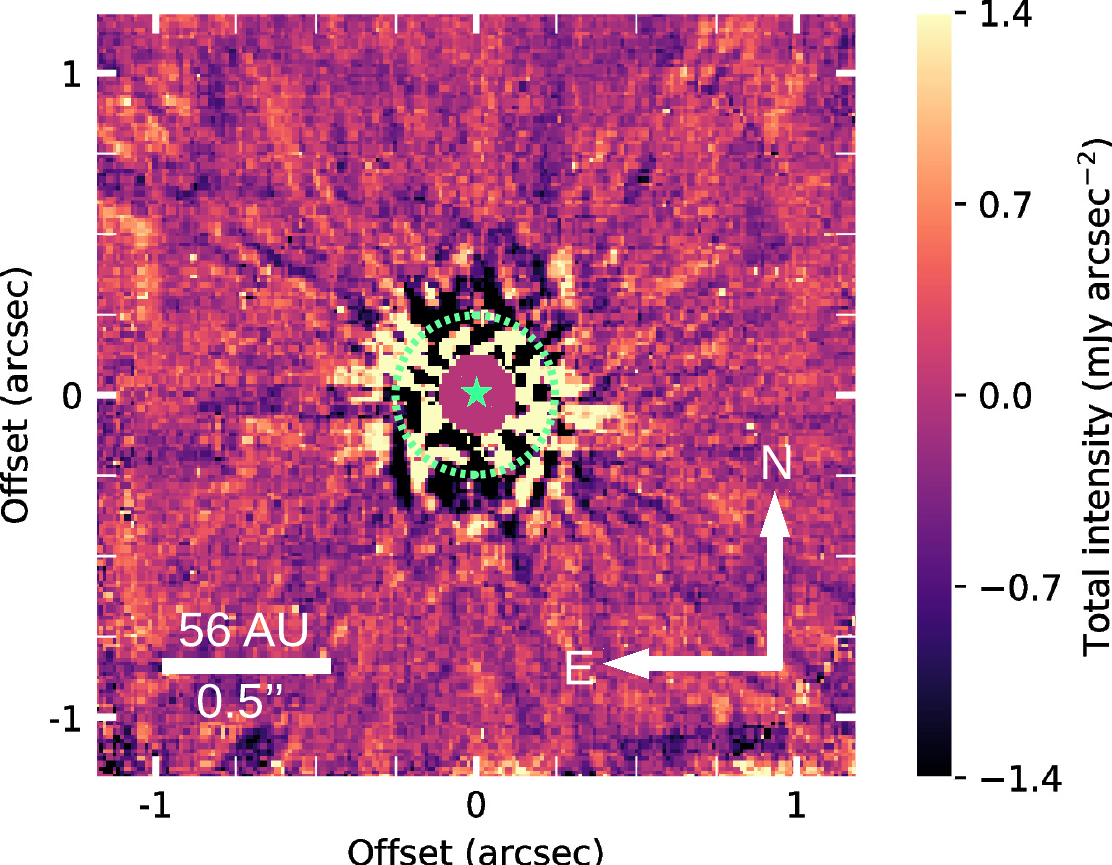}
  \caption{$H$-band total intensity (Stokes $I$) pyKLIP+ADI reduction of HD 141569A (Section \ref{stotintens}). Self-subtraction degrades the residuals, and there is no significant evidence of the presence of the inner ring. The image hints at the presence of the middle $220$ AU ring 1\arcsec to the east.}\label{JasonKLIP}
\end{figure}

We attempted to detect the total intensity emission from the inner ring in two different ways: from the combined polarization states in our PDI observations, and from the unpolarized IFS observations. Figure \ref{JasonKLIP} shows the result of the PSF subtraction of the Stokes $I$ image after applying \texttt{pyKLIP} \citep{Wang15KLIP} and Angular Differential Imaging \citep[ADI,][]{Marois06}.  Our shorter IFS sequence with GPI did not reveal neither the inner disk detected in Stokes $Q_{r}$ nor the previously known middle ring.

Our inability to detect the inner ring in total intensity is not entirely surprising. In extracting the Stokes $I$ signal from high-contrast observations we lose the differential imaging advantage of PDI. Moreover, the PSF subtraction process in total intensity also removes the smooth structure of the disk. Hence, we expect greater sensitivity to scattered light in our polarized light Stokes $Q_{r}$ image. In view of the low-SNR detection in total intensity we use only the polarimetric detection in Stokes $Q_{r}$ in the remainder of this study.

\section{Disk Modeling} \label{models}

\begin{table*}[ht]
\caption{Parameters probed in our exploration grid of models with MCFOST and best-fit values for the Stokes $Q_{r}$ image. \\
}\label{params}
\begin{tabular}{c@{}cc@{}c@{}cc}
& & & & &  \\
\cline{2-4}
\multicolumn{5}{c}{Fixed stellar parameters } \\
\cline{2-4}
&  Distance & $d(pc)$ & 111 & \\
&  Stellar Radius & $R_{*} (R_{\odot}$) & 1.66 R$_{\odot}$ & \\
&  Effective Temperature & $T_{\text{eff}} (K)$ &10000 & \\
&  Extinction & $E_{B-V}$& 0.144 & \\
&             & $R_{V}$   & 3.1 & \\

\cline{2-4}
\multicolumn{5}{c}{Disk model fixed parameters}\\
\cline{2-4}
& Inner Radius & $R_{in}$(AU) & 20 \\
& Outer Radius & $R_{out}$(AU) & 110\\
&Inner exponent & $\alpha_{\rm in}$ &  14\\
& Reference Radius & $R_{0}$(AU) & 45\\
& Solid material dust density & $\rho_{\text{dust}}$ (g cm$^{-3}$) & 3.5 & \\
& Flaring exponent & $\beta$ & 1 &  \\
& Vertical exponent & $\gamma$ & 1 &  \\
\cline{2-4}
& & & & &\\
\hline
\hline
Disk model free parameters &  & Sampling &Range &Number of values & Best Fit Value \\
\hline
  Inclination            & $i(^{\circ}$) & lin. in cosine & [40, 80] & 10 & $60^{\circ}\pm10^{\circ}$ \\
  Position Angle         & $PA(^\circ)$  & lin. & [$-25$, 15] & 9 & $5^{\circ}\pm 10^{\circ}$ \\
    Reference Scale Height & $H_{0}$ (AU)& lin. & [2.2, 36]& 9 &  $14_{-5}^{+9}$ \\
    Dust mass & $M_{d}$ (M$_{\odot}$)& lin. & [0.2, 2.0]$\times 10^{-6}$& 10 & $1.0\pm 0.4 \times10^{-6}$\\
    Outer exponent & $\alpha_{\rm out}$ & lin.& [$-3.5$, $0.0$] & 8 & $-1.0_{-1.0}^{+0.5}$ \\
    $R_{c}$ & $R_{c}$ (AU)& lin. & [24, 64] & 11 & $44_{-12}^{+8}$ \\
    Minimum grain size & $a_{\rm min}$ ($\mu$m)& log. & [0.5, 16] & 6 & $4_{-2}^{+4}$ \\
    Porosity & $p$ (\%) & lin.\ &[0, 20, 40, 60, 80] & 5 & 0 \\
    \hline
    Minimum reduced $\chi^{2}$ & & & & & 0.93\\
    \hline
\end{tabular}
\end{table*}
We model the resolved polarized Stokes $Q_{r}$ image of the inner disk with a radiative transfer model
to determine the disk geometry (Section \ref{model}) and dust properties (Section \ref{sec_model_estimates}). 
The same modeling tool can predict the SED for the disk models, but since the SED
is dominated by emission from dust outside the GPI field of view, we only compute model SED 
as a test to check against gross inconsistency (Section \ref{sec-sed}).

\subsection{Parameterization of the Dust Model}
\label{model}
We use the Monte-Carlo continuum radiative transfer and ray-tracing code MCFOST \citep{Pinte06, Pinte09} to compute synthetic observations of the SED and the Stokes $Q_{r}$ images of the disk around HD 141569A. MCFOST computes the scattering, absorption and re-emission events by dust grains by propagating photon packets throughout a cylindrical spatial grid in accordance with Mie theory. 

Dust grains are assumed to be in radiative equilibrium embedded in the radiation field from the host star. The sampling of our synthetic images is defined to cover the entire field of view of observations using GPI's pixel scale of $14.166 \pm 0.007$ mas lenslet$^{-1}$ \citep{DeRosa15}. The star is located at the center of the computational grid and the disk is centered on the star. To obtain the stellar luminosity, we fit the optical to near-IR SED of HD 141569A \citep{Tycho2, Mendigutia12, Cutri03}, keeping the stellar radius and foreground extinction as variables. We assumed a fixed distance of 111 pc 
with an effective temperature of $10000$ K for the star \citep{Merin04} and used NextGen photospheric models from \cite{NextGen99}. We obtain $L=25.48$ L$_{\odot}$ for the stellar luminosity. It agrees within uncertainties with the previous estimate of $25.77_{+1.63}^{-2.2}$ L$_{\odot}$ by \cite{Merin04} .

The disk geometry in cylindrical coordinates is described by the radial extent of the disk, with inner and outer radii $R_{in}$ and $R_{out}$, inclination $i$, position angle $PA$ and dust density distribution $\rho(r,z)=\Sigma(r)Z(r,z)$.  Following \cite{Augereau99a}, the dust distribution in the vertical direction is parameterized  within the MCFOST framework as an exponential with a shape parameter $\gamma$:
\begin{equation}
  Z(r,z)=\exp\left(-\frac{|z|}{H(r)}\right)^{\gamma},
 \end{equation}
 where the scale height is defined as $H(r)=H_{0}\left(\frac{r}{R_{0}}\right)^{\beta}$ at a fixed reference radius $R_{0}$=45 AU, and $\beta$ is a disk flaring parameter. Radially, the dust distribution follows a smooth combination of two power laws: 
\begin{equation}
 \Sigma(r) \propto \bigg\{\left(\frac{r}{R_{c}}\right)^{-2\alpha_{\text{in}}} + \left(\frac{r}{R_{c}}\right)^{-2\alpha_{\text{out}}}\bigg\}^{-1/2},
\end{equation}
 where $\alpha_{\text{in}} > 0$, $\alpha_{\text{out}} < 0$, and $R_{c}$ is the radial distance of the peak density of the grain distribution. The surface density of dust grains is represented as:
 \begin{equation}
 \sigma(r)=\int_{-\infty}^{+\infty}\rho(r,z)dz=\sigma_{0}\times \Sigma(r)\left(\frac{r}{R_{0}}\right)^{\beta},
 \end{equation}
 where $\sigma_{0}=2H_{0}\rho(R_{0},0)$ for $\gamma=1$. The maximum of the surface density is not at $R_{c}$, but at 
 \begin{equation}
r_{\text{max}(\sigma)} = \left(\frac{\Gamma_{in}}{-\Gamma_{out}}\right)^{\left(2\Gamma_{in}-2\Gamma_{out}\right)^{-1}}R_{c},\label{rmax}
\end{equation}       
where $\Gamma_{1}=\alpha_{\text{in}}+\beta$ and $\Gamma_{2}=\alpha_{\text{out}}+\beta$, which in our case is nearly identical to $R_{c}$.

We assume a disk with a constant opening angle, thus no flaring ($\beta = 1$), and adopt $\gamma=1$. Without better initial constrains on these parameters our assumption is reasonable for an optically-thin debris disk.

The surface brightness and thermal flux of the disk is controlled by the total mass in grains $M_{d}$ in the disk. To characterize the dust content in the disk, we adopt a power-law grain size distribution $dN(a) \propto a^{-n}da$ with $n=3.5$ as is commonly assumed for debris disks in a steady-state collisional cascade \citep{Dohnanyi69}. 
The size distribution has limits $a_{\text{min}}$ and $a_{\text{max}}$ and grain porosity $p$.
We fix $a_{\text{max}}=1$ mm and leave $a_{\text{min}}$ as a free parameter in the model. Within the MCFOST framework, the refractive index of porous grains is approximated from a mixture of solid grains with void particles following the Bruggeman mixing rule. Modeling is limited to a stellocentric disk populated by a single dust grain composition of amorphous magnesium-rich olivine with a dust grain density $\rho_{\text{dust}}=3.5$ g cm$^{-3}$. We adopted this dust grain composition to match the composition used by \citet{Thi14} and \citet{Mawet17}.

Our disk model parameterization thus comprises: $R_{in}$, $R_{out}$, $i$, $PA$, $H_{0}$, $M_{d}$, $\alpha_{\rm in}$, $\alpha_{\rm out}$, $R_{c}$, $a_{\rm min}$ and $p$. To reduce the burden of an 11 dimensional parameter exploration, we fixed $R_{in}$ and $R_{out}$ at 20 AU and 110 AU respectively. These disk boundaries are motivated by our inability to detect the inner disk within $\sim 0\farcs25$ and beyond 1$\arcsec$ from the star (Figure \ref{snrmap}). We also set $\alpha_{\rm in}=14$ as motivated by preliminary modeling that pointed to high $\alpha_{\rm in}$ values. Such a steep profile implies a sharp drop in density inside of $r_{\rm max}$. The probability density distribution for $\alpha_{\rm in}$ in those models plateaued at $\alpha_{\rm in} \geq 14$. 

By exploring each of the variable parameters at 5 to 11 discrete values, we construct a model grid with over $3\times 10^{7}$ grid points for the remaining eight free parameters. Table \ref{params} details the full set of parameters involved in the modeling, including additional fixed parameters for the star and disk.  

\subsection{Polarized Scattered Light Modeling and Best Fit Estimates}
\label{sec_model_estimates}

\begin{figure*}[ht!]
 \center
 \includegraphics[width=7.5in]{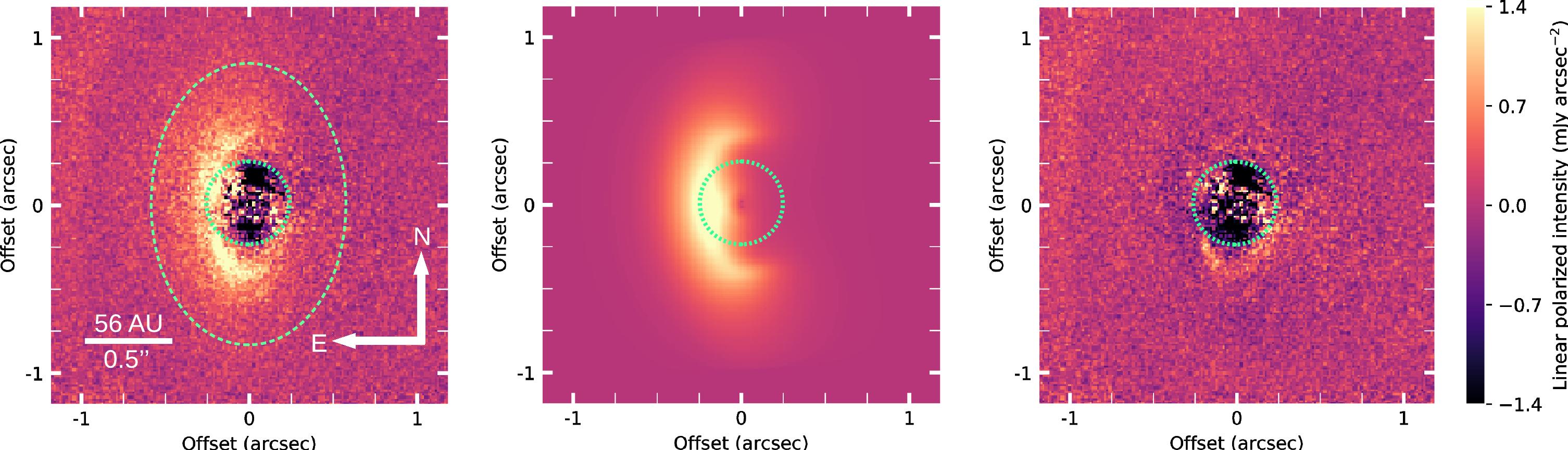}
  \caption{Modeling of the inner HD~141569A dust ring with MCFOST. The central $0\farcs25$ circular region is the same as in Figures~\ref{obsGPI}--\ref{JasonKLIP}, and is ignored in the modeling. {\it Left:} Observed Stokes $Q_{r}$ image with the fitting region delimited between the circular and elliptical dashed lines. {\it Center:} Stokes $Q_{r}$ image from best-fit model. {\it Right:} Residuals of the Stokes $Q_{r}$ image minus the best-fit model. A residual arc-like polarized emission is seen to the south. We have used the same intensity scale and color stretch as for the Stokes $U_{r}$ image in Figure~\ref{obsGPI}.}\label{mcfostmodel}
\end{figure*}

In preparation for the fitting procedure, we first binned the images in $3\times 3$ pixels to reduce the effects of correlated errors. GPI's resolution element is about 3 pixels in the $H$ band, and therefore, the binning should not lead to loss of spatial information. Following \cite{Max15}, we used the $3\times 3$ binned Stokes $U_{r}$ to derive the uncertainties used to fit the binned Stokes $Q_{r}$ image. For each position in the binned Stokes $Q_{r}$ image, the uncertainty is calculated as the standard deviation of a 3 pixel-wide annulus on the binned $U_{r}$ image. These steps return the uncertainty map $\sigma$ to use in the $\chi^{2}$ minimization procedure:
\begin{equation}
\chi^{2}=\Sigma_{i}^{Npix}\left(\frac{Obs_{i} - Mod_{i}}{\sigma_{i}} \right)^{2} . \label{chichi}
\end{equation}
The fitting occurs within an $0\farcs85 \times 0\farcs61$ (radius) elliptical region centered on the star and excludes the central $0\farcs25$-radius circular region to avoid PSF subtraction residuals (Figure \ref{mcfostmodel}, left). The fitting region in the binned image has 767 resolution elements. We opt for an elliptical fitting region rather than a circular one because Poisson noise from the disk rather than the stellar halo is the main limiting factor for our sensitivity. 

The outcome of our modeling strategy is presented in Figure \ref{mcfostmodel} which shows the Stokes $Q_{r}$ image (left), the best-fit model (center), and the residuals (right): all displayed on the same color and intensity scale as the Stokes $U_{r}$ image in Figure \ref{obsGPI}. Our best-fit model returns a reduced $\chi_{r}^{2}=0.93$ and so provides a good match to the Stokes $Q_{r}$ image. Our best-fit model Stokes $U_{r}$ image contains very little flux, on the order of $0.1\%$ of the model Stokes $Q_{r}$. This indicates that the disk is optically thin with an inferred mid-plane optical depth of $\tau_{1.65\mu m} = 10^{-2}$ from MCFOST.

\begin{figure*}
\begin{center}
\includegraphics[width=6.0in]{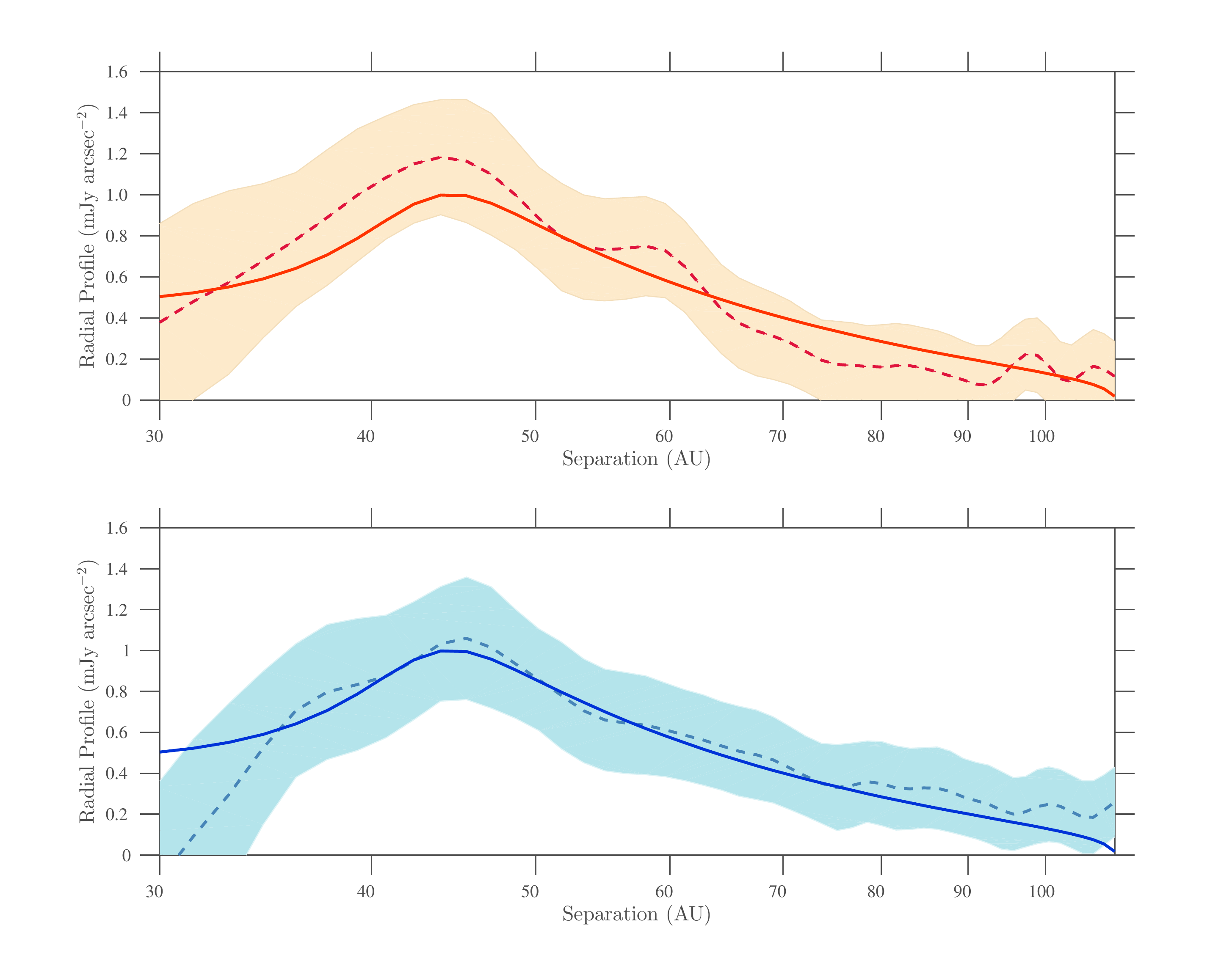}
\end{center}
\caption{Radial profiles along the semi-major axis ($PA= 5^{\circ}$) of the Stokes $Q_{r}$ image (dashed line) and of the best-fit models with MCFOST (solid line) inwards to 30 AU ($0\farcs27$):\emph{ north (top) and south (bottom).} Shaded areas indicate the dispersion calculated using the uncertainty map derived from the Stokes $U_{r}$ image (Figure \ref{obsGPI}, right panel). The best-fit model falls within uncertainties and in general agrees with the observed profiles.} \label{profiles}
\end{figure*}

As a consistency check on our best-fit solution, we plot radial profiles of the polarized emission along the semi-major axis of the disk, and compare them to the prediction from the model (Figure~\ref{profiles}). We see that the model follows the radial profile well, and stays within the 1-$\sigma$ uncertainty band at most separations, except inwards of 31 AU to the north, where it overestimates the observed emission. There is also residual emission to the south around $PA=150^\circ$ that has no counterpart to the north (Figure \ref{mcfostmodel}, right), and that we discuss in Section \ref{discussion1} and \ref{discussionspiral}.

\begin{figure*}[!ht]
\begin{center}
\includegraphics[width=10in]{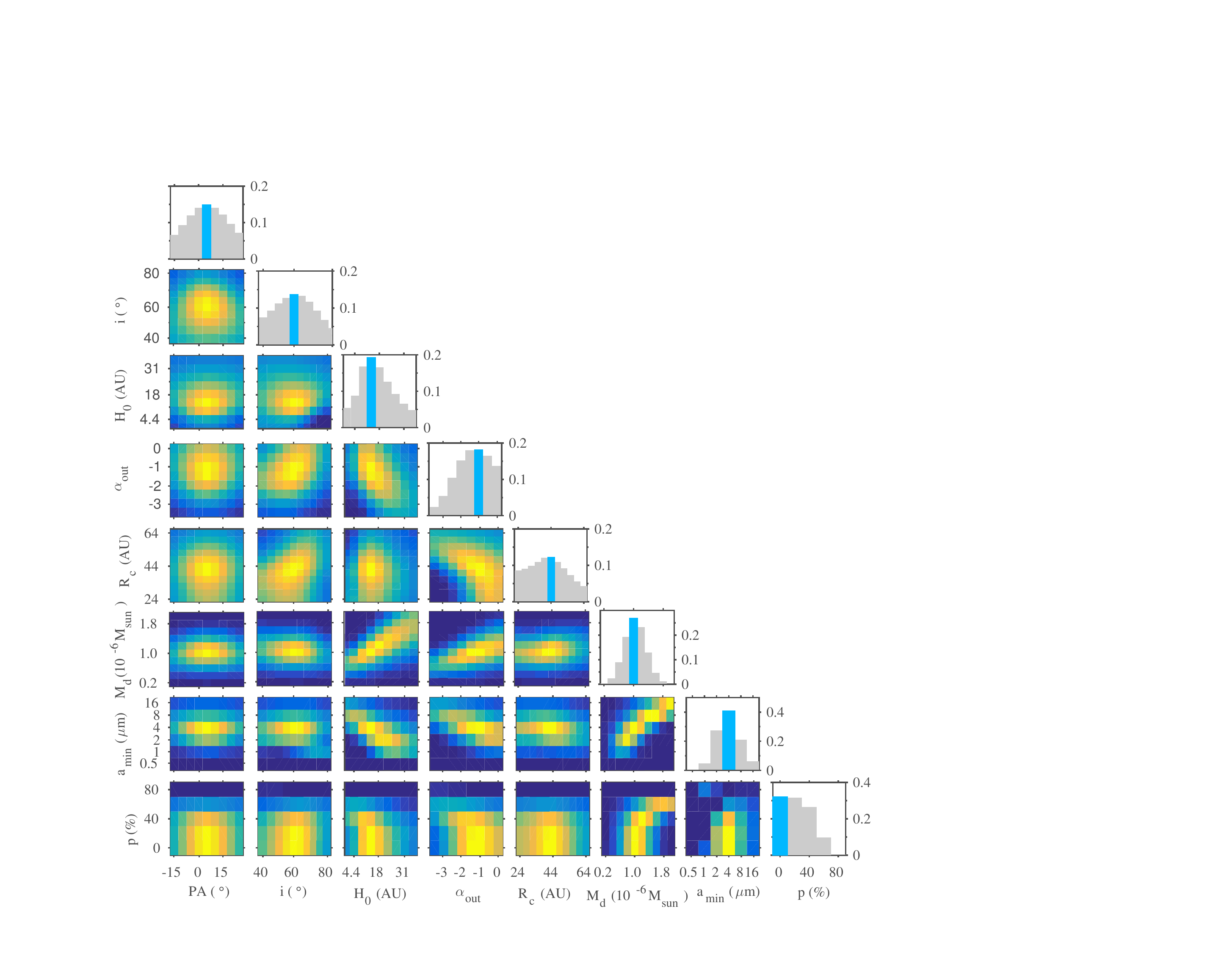}
\end{center}
\caption{Normalized probability density distributions for each parameter in the grid of models explored to fit the Stokes $Q_{r}$ image with MCFOST. Brighter yellow regions in the color maps correspond to higher probability densities. A blue bar in each of the histogram panels along the diagonal shows the best-fit value for each parameter.}\label{corner}
\end{figure*}

The overall consistency of the model and image radial profiles give us confidence that we have an adequate understanding of the dust disk parameters. Assuming that our observational errors are approximately Gaussian, and adopting a flat prior for each parameter, the Bayesian probability of our model given the data is:
\begin{equation}
P \propto \exp{\left(-\frac{\chi^{2}}{2}\right)}.
\end{equation}
To estimate the probability density distribution for any parameter, we marginalize $P$ over the remaining parameters as shown in Figure \ref{corner}. 

We obtain good constraints for $\alpha_{\text{out}}$, $M$ and $a_{\text{min}}$, whose probability density distributions are approximately Gaussian. Unfortunately, our modeling returns poor constrains on the viewing geometry of the disk and on some of the parameters that describe the spatial distribution of the dust. The posterior distributions for $i$ and $PA$ in Figure \ref{corner} are broader than we anticipated. Hence, we cannot establish a proper uncertainty. This suggests that polarized intensity alone is not adequate for determining the geometry, since only the Eastern half of the disk is detected. In addition, for $H_{0}$ and $R_{c}$ our modeling returns non-symmetric skewed distributions. We use the 68\% confidence intervals as estimates for the uncertainties on the best-fit parameter values for $i$, $PA$, $H_{0}$, $\alpha_{\rm out}$, $R_{c}$ and $M_{d}$. For $a_{\rm min}$ and $p$, we use the 90\% confidence interval instead, as motivated by our coarser sampling in these parameters.

Our best-fit disk model has $i \sim 60^{\circ}$, position angle $PA \sim 5^{\circ}$, scale radius $R_{c}=44_{-12}^{+8}$ AU, a rather large reference scale height $H_{0} = 14_{-5}^{+9}$ AU (which at a reference radius of $R_{0}=45$ AU gives an opening angle of 17$^{\circ}$), and a shallow outer exponent $\alpha_{\text{out}}=-1.0_{-1.0}^{+0.5}$. 
The total dust mass of $1.0\pm 0.4 \times 10^{-6}$M$_{\odot}$ is within the range $(0.03-1) \times 10^{-5}$ M$_{\odot}$ estimated from SED modeling \citep{Zuckerman95, Sylvester96, Merin04, Thi14}. The best-fit minimum grain size is $a_{\text{min}}=4_{-2}^{+4}$ $\mu$m (90\% confidence interval). The minimum grain size agrees with the blow-out grain radius  \citep{Burns79} in a gas-poor disk, with $r_{\text{blow-out}}=4.2 \mu$m for spherical silicate grains of density $\rho=3.5$ g cm$^{-3}$, assuming a radiation pressure efficiency factor of $Q_{p}=1$. We discuss the blow-out radius with respect to the gas content of the inner disk further in Section \ref{sec_amin}. The best-fit model favors a population of solid dust grains, porosity $p=0\%$, although it is consistent with porosity up to $p=40\%$ within the 90\% confidence limit. 

As seen in Figure~\ref{corner}, the total dust mass $M_d$ is strongly degenerate with several disk parameters, notably the minimum grain size $a_{\rm min}$, the outer exponent $\alpha_{\rm out}$ and, surprisingly, the disk scale height $H_0$. There is no evident mechanism by which the scale height would be degenerate with the disk mass for an optically thin disk. A third parameter could set this correlation but the relationship remains unclear. All of these degeneracies preclude us from placing fully independent constraints on $M_d, a_{\rm min}, H_0$, or $\alpha_{\rm out}$.  In Section~\ref{sec_amin} we discuss an independent and more stringent constraint on the minimum grain size $a_{\text{min}}$, arising from the lack of measurable signal in the Stokes $U_{r}$ image.

\subsection{Comparison of the Scattered Light Model to the SED} \label{sec-sed}
\begin{figure}[htbp]
\hspace*{-1.0cm}
  \includegraphics[width=3.9in]{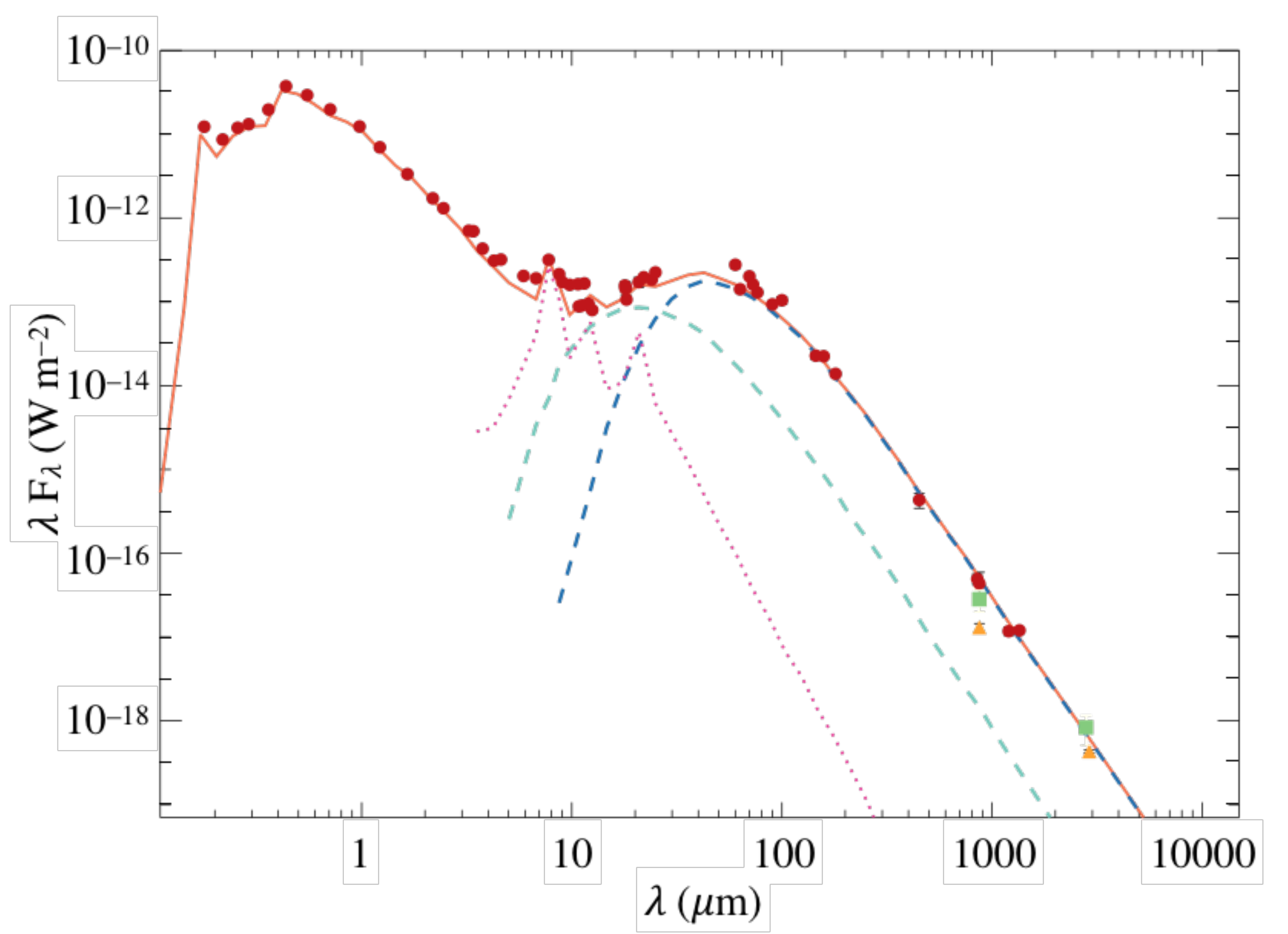}
 \caption{Measured data points and predicted MCFOST SEDs. Observations comprise the photometric compendium by \cite{Thi14} and references therein (filled circles), SMA photometry by \citet[][filled squares]{Flaherty16}, and ALMA photometry by \citet[][filled triangles]{White18}. The model SEDs represent: the best-fit model to the Stokes $Q_{r}$ image of the HD 141569A inner disk (dark blue dashed line), a fourth innermost disk (light blue dashed line), and PAH emission (magenta dotted line). The total emission from these components and the star is shown with a solid line.}
\label{sedmodel}
\end{figure}

We compare the predicted thermal emission from our best-fit model to the Stokes $Q_{r}$ image against the SED of HD~141569A in Figure \ref{sedmodel}. Fitting the SED is not a part of our search for the model that best matches the scattered light emission. It is merely a consistency check on our scattered light modeling assuming that the same dust population is responsible for both disk tracers. 

We use the compendium of photometric data from \cite{Thi14} with updated photometry in the optical \citep{Tycho2, Mendigutia12}. We expand this dataset by including recent sub-millimeter and millimeter photometry with ALMA \citep{White18} and with the Submillimeter Array \citep[SMA;][]{Flaherty16}. We note that the photometric measurements in \cite{Thi14}, \cite{Flaherty16} and \cite{White18} come from instruments with different resolutions and beam sizes.  These range from FWHW $\approx1\arcsec$ seeing-limited optical/near-IR measurements, to a beam of $5\farcs1 \times 4\farcs2$ for the 2.8 mm SMA photometry, and up to $2\arcmin\times5\arcmin$ for 60$\micron$ photometry from the Infrared Astronomical Satellite (IRAS).

We also include a PAH component, motivated by resolved (FWHM = 0.26\arcsec) observations with the VLT Imager and Spectrometer for the mid-IR \citep[VISIR;][]{lagage04} in the PAH1 filter ($\lambda_{c}=8.6\mu$m, $\Delta\lambda=0.42$). These reveal a disk out to $1\arcsec$ along the semi-major axis \citep{Thi14}. PAHs in the circumstellar environment of HD 141569A are responsible for the emission features in the mid-IR at 7.7, 8.7, 11.3 and 12.7$\mu$m \citep{Sylvester96}. We include the PAH component (magenta dotted line) with the sole goal of approximating the emission features in the mid-IR modeled in previous studies \citep{Li03, Thi14}. PAHs were added with the adopted single molecule size of 6.84 \AA{} and fixed mass of $1.6\times10^{-10}$ M$_{\odot}$ from \cite{Thi14}.

The thermal emission from our scattered light model of the 44~AU disk reproduces the observed $\geq$50$\micron$ fluxes adequately.  The emission peaks at a wavelength of 45$\micron$ (and so has a characteristic dust temperature of 67~K) and a flux of $1.6\times10^{-13}$ W~m$^{-2}$.  This is within factors of 1.0--1.8 of the observed 60$\micron$--100$\micron$ far-IR fluxes measured with IRAS, Spitzer, and Herschel \citep[][and references therein]{Thi14}.  The range is entirely due to the scatter in the flux measurements from different instruments.

At $>$140$\micron$ wavelengths the predicted thermal emission from the 44~AU disk matches the SED unexpectedly well, considering that we did not include any SED information in our modeling. The published far-IR and millimeter flux measurements are obtained from much wider beams that incorporate the outer two debris rings, so we expect that the fluxes should be higher than the predicted brightness of the 44~AU ring.  While an SED analysis of HD~141569A is relegated to a future study, there are two independent lines of evidence that suggest that the 44~AU ring dominates the $>$100$\micron$ thermal emission.  First, the emission resolved in the ALMA 870$\micron$ continuum and CO maps in \citet{White16} is consistent with origin in a $\sim$50 AU ring: likely the same dust ring as resolved by GPI.  ALMA shows no evidence of significant millimeter emission associated with the middle (220~AU) dust ring.  Second, with the middle dust ring being $\sim$5 times wider than the inner 44~AU ring, its characteristic dust temperature will be $\sim$$\sqrt{5}$ cooler, so $\sim$30~K.  A strong 30~K dust component will produce a notable bump in the SED around 100$\micron$. However, the observed $\geq$60$\micron$ fluxes closely follow a Rayleigh-Jeans distribution.  Hence, we believe that the 44~AU dust disk resolved by GPI accounts for most of the long-wavelength flux from the HD~141569A circumstellar disk.

Finally, we note that even while it reproduces the observed $\geq 60\micron$ fluxes adequately, the combined SED of our scattered light and PAH emission model is underluminous between 8--30\micron. A fourth component, interior to the one seen by GPI, is likely present around the star. To correct for this flux deficit in the mid-IR, we model an innermost 5--15 AU dust disk (light blue dashed line in Figure \ref{sedmodel}) in MCFOST. We discuss this innermost disk further in Section \ref{sec_fourth_ring}. 




\subsection{The Effect of Mie Scattering Assumptions on Disk Modeling Results}
\label{mie_caveat}

The synthetic models in our modeling procedure rely on Mie scattering theory to compute the scattering, opacity and absorption cross sections of dust grains in the circumstellar environment of HD 141569A. Although Mie theory has been extensively used in numerous debris disk studies, it has limitations.

In Mie scattering, individual dust grains are idealized as uniform solid spheres, an assumption likely not applicable to dust grains growing through agglomeration in debris disks. Hence, Mie theory could struggle to accurately predict the light scattering and thermal processes in debris disks.  More elaborate models with irregularly shaped dust grain aggregates indicate that Mie theory does not reproduce well the scattering phase function at angles $\theta > 90^{\circ}$ \citep{Min16}.  Accordingly, \citet{Milli17} show that Mie models fail to reproduce the scattering phase function for the HR 4796A dusty debris ring.  Such discrepancies likely bias the determination of certain model parameters, in particular the viewing geometry, minimum grain size, dust composition, dust mass, and porosity.

While we have adopted Mie theory-based models for our analysis for the sake of comparison with previous radiative transfer modeling on the dust-scattered light from the HD~141569A debris disk \citep{Jonkheid06, Thi14, Mawet17}, we note the above limitations ahead of the ensuing discussion.

\section{Discussion}\label{discussion}

\subsection{Morphology of the HD 141569A Inner Disk in Polarized Scattered Light: Comparison to Previous Observations }\label{discussion1}

PDI with GPI has revealed the clearest view of the inner disk around HD 141569A inwards to an inner working angle (IWA) of $0\farcs25$ (28 AU). The IWA achieved with GPI shows the 44 AU ring morphology that supersedes any lower-SNR detections from images with larger IWAs. 

The highest SNR obtained on the 44 AU ring previously in scattered light is through $L_{\text{p}}$-band AO and vortex coronagraphy with Keck/NIRC2 (FWHM $\approx 0\farcs08$, effective IWA $\simeq$ 0\farcs16) by \cite{Mawet17}, and also reported with Keck/NIRC2 in $L_{\text{p}}$-band by \cite{Currie16} down to 0\farcs25. \cite{Perrot16} report even higher angular resolution $H$-band AO observations with VLT/SPHERE (FWHM $\approx 0\farcs040$, IWA=0\farcs093), as do \cite{Konishi16} with HST/STIS (FWHM=0\farcs04, IWA=0\farcs40). Compared to our PDI with GPI, all of these previous non-polarimetric observations have lower SNR because of the inability to employ simultaneous differential imaging through polarimetry, but relying solely on ADI instead. Diffuse structures with significant axial symmetry, such as debris disk seen at low to moderate inclinations, are challenging to extract with ADI/KLIP. PDI with GPI has produced a high-SNR detection of the disk with much lower reduction-dependent systematics. 

Our GPI polarized light image confirms the clearing within the 44 AU ring first reported by \cite{Perrot16}. \cite{Thi14}, \cite{Konishi16} and \cite{Mawet17} report PAH emission or scattered light emission from dust at similar separations, but describe its radial dependence with a single-exponent power law that decreases with separation. As seen in Figure \ref{profiles}, we clearly resolve the peak at 44 AU that is well modeled with exponential drop-offs on either side. The width of the
ring in polarized light (FWHM $\sim$ 30 AU) is greater than reported (FWHM $\sim$ 10 AU) from the unpolarized VLT/SPHERE observation of \cite{Perrot16}, which we attribute to self-subtraction in the various KLIP reductions of the SPHERE images.

Consistent with \cite{Perrot16} and \cite{Mawet17}, we find a north-south asymmetry in the brightness of the 44 AU ring, which we now reveal as a high-SNR arc-like structure to the south. We do not see evidence of the other clumps reported in these studies, and suspect they may be related to image artifacts. 

The viewing geometry of our best-fit model, $i\sim60^\circ$ and $PA\sim5^\circ$, is similar to previous findings from scattered light total intensity observations at lower SNR. \cite{Perrot16} report that the inner ring has an inclination of $i = 57.9^{\circ}\pm 1.3^{\circ}$ and a $PA$ of $353.7^{\circ}\pm1.1^{\circ}$ from observations with VLT/SPHERE, while \cite{Mawet17} report $i = 53\pm 6^{\circ}$ and $PA = 349\pm 8^{\circ}$ from $L_{\text{p}}$ observations with Keck/NIRC2.
However, unlike all previous direct imaging (non-polarimetric) observations, which point to a relatively well-constrained slightly W of due N orientation of the semi-major axis of the inner ring, our polarimetric observation produces a less well constrained, slightly E of due N PA. Our values for $i$ and $PA$ follow broad distributions in Figure~\ref{corner} that suggest uncertainties of about 10$^\circ$. Our own lower-SNR $H$-band total intensity image (Figure~\ref{JasonKLIP}) suggests a slightly clockwise ($PA\sim 350^\circ$) orientation of the scattered light emission: in agreement with \citet{Perrot16} and \citet{Mawet17}, and contrary to our polarized image. Effectively, the orientation of the inner disk from previous total intensity observations is in good alignment with the outer disk, whereas the PDI observations are less conclusive.

It is at first surprising that our polarimetric image does not produce better constraints on the geometry. The SNR of all of our own and of the previously published total intensity observations is much lower.  The reduced total intensity images also suffer from the typical PSF-subtraction systematics for extended emission around bright stars: the result of ADI mode observations and image reduction with the KLIP algorithm. Thus, it is possible that the uncertainties in \cite{Perrot16} have been underestimated, and our values are closer to agreement with theirs because of larger uncertainties \citep[as in][]{Mawet17}.  Nonetheless, the discrepancy between the polarized and the total intensity geometry is still unusual, as is the inability to get better geometric constraints from our polarized intensity images. 

We suspect that the failure of our modeling to produce better viewing geometry constraints may be a consequence of detecting only half of a radially extended disk in polarized intensity. The detection of this non-axi-symmetric half disk, combined with the broader arc-like feature to the south (\S~\ref{sec_model_estimates}, Figure \ref{mcfostmodel}, right), may favor models with $PA$'s flipped around the north-south axis. In addition, as discussed in Section~\ref{mie_caveat}, it is possible that the scattering phase function derived in our Mie-based radiative transfer MCFOST model may not agree well with the behavior of debris disk dust grains. The detected side of the debris ring is seen at a $>$90$^\circ$ scattering phase angle, and as in the case of the HR~4796A debris ring \citep{Milli17}, the polarization and intensity of the scattered light may not be well represented by Mie theory.

\subsection{The Arc-like Structure: a Spiral Arm?}\label{discussionspiral}
\label{planet}

The arc-like asymmetry spans between $120^{\circ}-170^{\circ}$ in $PA$. This feature is not an outcome of the data reduction process because the PDI image requires no further PSF subtraction, but is rather a true brightness enhancement in the ring. A disk with a stellocentric offset could offer a possible explanation for the enhanced emission from the arc-like structure to the south. The portion of the disk closer to the star would appear brighter than the other side, leading to pericenter glow \citep{Wyatt99}. Our modeling procedure did not include stellocentric offsets, and so we cannot check for pericenter glow. \cite{Perrot16} find that the inner ring has a stellocentric offset of $15.4\pm 3.4$~mas ($1.7\pm0.4$~AU)  to the west, but only a negligible one, $1.2\pm 9.4$~mas ($0.1\pm1.1$~AU), to the north. Therefore a north-south stellocentric offset is not the likely cause of the brightness enhancement to the south.

\begin{figure*}
\begin{center}
\includegraphics[width=6.2in]{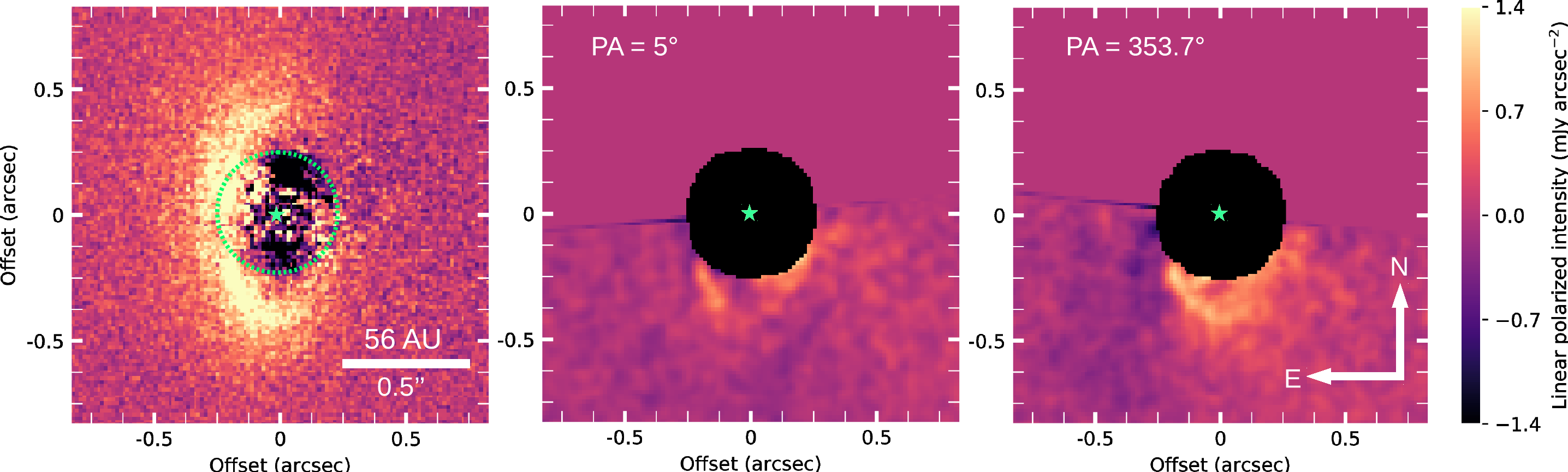}
\end{center}
  \caption{Revealing the southern arc on the 44~AU dust ring around HD~141569A by mirroring the northern part of the disk and subtracting it from the southern part. {\it Left:} The polarized intensity image from GPI. {\it Middle:} Mirroring and subtraction, assuming the best-fit geometry of the model of the GPI polarized light $H$-band emission. The arc extends over 120--170$\degr$ and contributes $\sim$40\% of the total disk brightness level. {\it Right:} Mirroring and subtraction, assuming the inferred geometry of the SPHERE total intensity $H$-band emission.  The arc can be traced counter-clockwise to at least 190$\degr$ at 40--50\% of the overall disk brightness.}\label{fig_mirrored_subtractions}
\end{figure*}

For a clearer view of the morphology of the asymmetry, we mirror and subtract the northern portion of the disk from the southern half. Given the ambiguity of the disk's orientation, we perform two different subtractions. In the first case we mirror around the semi-minor axis of the best-fit model of the polarized emission seen with GPI; i.e., the $PA$ of the semi-major axis is 5$\degr$ (Figure~\ref{fig_mirrored_subtractions}, middle panel). In the second case we use the geometry inferred from the total intensity image from SPHERE, with semi-major axis $PA$ of 353.7$\degr$ (Figure~\ref{fig_mirrored_subtractions}, right panel). Both subtractions show the excess emission to the south as a remnant arc at SNR levels of 4--5 per GPI pixel along the peak of the emission. The residuals in the $PA=5\degr$ case are closer to zero, which is why our modeling of the polarized light emission prefers that geometry.  In this case the arc contributes up to 40\% of the surface brightness of the debris disk between $120\degr<PA<170\degr$, and has a net integrated surface brightness of $\approx$5~mJy within a 10~AU-wide region centered on the peak emission at 44~AU.  In the semi-major axis $PA=353.7\degr$ case the residuals are more uniform, and also more positive.  The arc contributes 50\% of the disk surface brightness at $PA=130\degr$, and wraps counter-clockwise to at least $PA=190\degr$, where it is still at $\sim$40\% of the disk's total brightness before the signal diffuses into the residual noise.  The net integrated surface brightness of the arc in the same $120\degr<PA<170\degr$ region is $\sim$15~mJy, and 25~mJy if extended up to $PA=190\degr$.

Without a higher-SNR total intensity image of the disk, we can not decide in favor of one disk PA vs.\ the other.  However, both point to the existence of an arc on the inner ring of the HD~141569A disk that contributes between 40\%--50\% of the total disk flux to the south-east. Such arc-like structures are known on the outer two rings of HD~141569A \citep{Mouillet01, Clampin03, Perrot16, Mawet17}. We believe this to be the first convincing detection of such an arc on the inner ring. The location and extent of the feature match the observed brightness enhancement in the SE section of the inner ring in \cite{Perrot16} and the enhancement in the CO zeroth moment map in \citet{White16}. Our best-fit model indicates an average temperature of $\sim 90$ K at the location of the arc, well above the sublimation point of CO. The CO production mechanism may thus be linked to the dust over-density: pointing to ongoing destruction of CO ice-rich planetesimals. The destruction cascade itself may be triggered by an unseen body that is also responsible for producing the arc: as a spiral arm.

Spiral arm structures have been discovered in near-IR scattered light imaging observations of a few circumstellar disks (e.g., AB Aur, \citealt{hashimoto11}; HD 142527, \citealt{avenhaus14}; SAO 206462, \citealt{muto12}; MWC 758, \citealt{grady13}; HD 100453, \citealt{wagner15hd100453}). Two mechanisms are capable of driving such arms in gas-rich protoplanetary (and transition) disks: gravitational instability \citep[e.g.,][]{dong15giarm}, and interaction between the disk and a planetary or stellar companion \citep[e.g.,][]{dong15spiralarm, dong16hd100453}. In optically thin debris disks with much lower gas-to-dust ratios, photoelectric instability \citep{klahr05, besla07} may also lead to clumping of dust into structures. Spiral density waves are one of the hypotheses invoked for explaining the radially moving dust enhancements in the edge-on AU Mic debris disk \citep{BoccalettiAUMIC15}. However, typically multiple rings and arcs (i.e., broken rings), instead of one or two spiral arms, are seen in simulations \citep{lyra13pei, richert18}. While the HD 141569A disk is almost certainly too low in mass to be gravitationally unstable, the possibility that the detected spiral-arm-like feature is driven by an unseen planet is exciting. \citet{dong16armviewing} showed that spiral arms driven by giant planets in disks at modest to high inclinations may appear very close to, or be part of, the disk ring sculpted by the planet. In particular, the 50$^\circ$ and 60$^\circ$ inclination panels in Figure 8 of \citet{dong16armviewing} show intriguing similarities with the HD 141569A inner-disk spiral arm in Fig. 7. The weak contrast of the arm in HD 141569A indicates that if it is planet-driven, the planet is most likely Jovian or smaller \citep{dong17spiralarm}.

\subsection{Disk Opening Angle}
\label{opening_angle}
While it is much less powerful a constraint than it is in (optically thick) protoplanetary disks, our modeling allows us to place approximate constraints on the opening angle or scale height of the resolved inner disk. The best-fit model indicates a rather large reference scale height of $H_{0}=14$ AU at the $R_{0}=45$ AU reference radius, so an opening angle of $17^{\circ}$. This is above expectations even for a transitional disk, although values as small as 10\% are within the 84\% confidence limit. With such a large disk opening angle, the best-fit model incorporates significant scattering at angles smaller than the $\sim$ $30^\circ$ expected from a perfectly flat disk (given inclination of $i\sim60^\circ$). However, if our Mie theory-derived scattering phase function is wrong, this constraint is not to be trusted.

Previous determinations of the opening angle range from 5\%--10\% \citep{Thi14} to 23\% \citep{Merin04}. The latter is from SED fitting alone, and while consistent with our finding, it is not very well constrained. The \citeauthor{Thi14} determination pertains to the gas disk geometry and is constrained from \textit{Herschel} measurements of the [\ion{C}{2}]/CO $J=3-2$ line flux ratio, which traces the efficiency of CO photodissociation as a function of gas scale height. However, without having spatially resolved the inner disk, \citeauthor{Thi14} adopt a model with a gas surface density peak at $\approx$28~AU (Fig.~6 in that paper), whereas we resolve the brightness peak of the inner dust disk at 44~AU (Fig.~\ref{profiles}). If the gas and the dust are well-mixed, as assumed by \citeauthor{Thi14}, then a cooler gas disk would require a greater scale height to produce the same [\ion{C}{2}]/CO $J=3-2$ line flux ratio. It is therefore likely that under the joint constraints from the \textit{Herschel} gas abundances and the GPI resolved dust disk morphology, the gas disk has a $>$10\% opening angle consistent with the wide dust disk opening angle found here.

\subsection{Independent Constraint on the Minimum Grain Size from Polarimetry}
\label{sec_amin}
Our model of the polarized scattered light (Section~\ref{sec_model_estimates}) produced a minimum grain size of $a_{\rm min}=4^{+4}_{-2}\micron$, consistent with the 4.2$\micron$ blow-out grain radius around HD~141569A. The result for the blow-out radius is meaningful as long as the inner disk remains gas-poor.  Conversely, in gas-rich disks with interstellar medium-like gas-to-dust ratios of $\sim$100, gas drag dominates the dynamics of small grains.

Based on a total gas mass estimate of 200$M_\oplus$ ($6\times10^{-4}M_\odot$), \citet{Thi14} find an average gas-to-dust ratio of 50--100 over the full extent of the $\sim$500~AU disk.  This would preclude a meaningful radiation-pressure estimate of the blow-out radius.  However, \citet{White16} find a much lower (H$_{2}$) gas mass of $1.5 M_{\oplus}$ within 210~AU, and observed that the inner $\sim$50~AU region shows only tenuous CO emission.  Hence, the inner disk that we resolved with GPI is substantially more deprived of gas.
Adopting the \citet{White16} gas mass, and our best-fit dust mass of $\sim10^{-6}M_\odot = 0.3M_{\oplus}$ for the 44 AU ring (Table~\ref{params}), yields a gas-to-dust ratio of $\sim$5.  The actual gas-to-dust ratio in the 44~AU ring is likely lower, since the \citet{White16} gas mass estimate encompasses both the inner 44~AU and parts of the middle 220~AU ring.  Therefore, we expect that the smallest grains in the inner ring are more strongly affected by radiation pressure than by gas drag.

Our best-fit value of $a_{\rm min}=4^{+4}_{-2}\micron$ for the minimum grain size is marginally consistent with previous findings from scattered light observations \citep{Marsh02, Mawet17} but differ from values inferred from SED modeling \citep{Thi14, Mawet17}. The resolved Keck II mid-IR observations of \citet{Marsh02} yield a best-fit ($\chi^2_\nu=1.23$) minimum grain size of 1--3$\micron$---in agreement with our findings---although fits with either ISM-sized 0.1$\mu$m ($\chi^2_\nu=1.40$) or large ($\gtrsim$30$\micron$) blackbody grains ($\chi^2_\nu=1.50$) are also satisfactory.  Using Mie scattering assumptions and MCFOST for modeling, \citet{Mawet17} find that a population of dust particles of pure olivine with $a_{\rm min}=10\mu$m provides the best fit to the resolved $L_{\text{p}}$-band scattered light emission between 20--90 AU. At the same time, \citeauthor{Mawet17} also find that a minimum grain size of 0.5$\micron$ best fits the SED, echoing the findings from SED model fitting by \cite{Thi14}. The combined best fit to the $L_{\text{p}}$ image and SED in \citet{Mawet17} yields an even smaller minimum grain size: $a_{\rm min}=0.1\micron$.

The preference for very small (0.1--0.5$\mu$m) grains in SED modeling points to the presence of a warm dust component that may not be well represented by an extrapolation of an index $n=3.5$ (collisionally-dominated) grain size distribution below 1 $\mu$m.  The collisional cascade may not be equally efficient at all grain sizes, or at all radial separations in the disk.  Such is the case at least for large grains around HD~141569A, as multi-band 0.9--9 mm ALMA and VLA observations show that at millimeter sizes the index is $n=2.95\pm0.1$ \citep{White18}.  

Our polarization observations are uniquely diagnostic of the presence of sub-micron grains because of their polarization properties. Specifically, scattered light models with a significant population of $a_{\rm min} < 0.8\mu$m are rejected because they produce  negative polarization in Stokes $Q_{r}$ that is not observed by GPI. We similarly rule out highly porous ($p>60\%$) materials. Thus, with the added power of near-IR polarimetry, we conclude that the population of $\leq$1.0$\mu$m grains in the 44~AU dust ring is not significant enough to be detectable in polarized light. A trace population may nonetheless exist, and could be responsible for the observed PAH emission. We use this result to argue for a fourth, innermost and unseen component of the HD~141569A debris disk in Section  \ref{sec_fourth_ring}.  

We again caution that this analysis is rooted in Mie theory, that may not yield the correct ratio of scattered to absorbed (and emitted) flux.  A non-Mie treatment, could yield different constraints on the minimum grain size from the observed polarization signatures.  However, an additional argument against the presence of a large reservoir of sub-micron grains is the lack of a 10$\micron$ silicate feature in the Spitzer IRS spectrum of HD~141569A \citep{Sloan_etal05}, as also argued by \citet{Thi14}.  Hence, we find that a warm ring of dust grains several microns in size and lying interior to the one resolved with GPI offers the simplest explanation for the extra 8--30$\micron$ emission from HD~141569A.


\subsection{An Unseen Innermost (Fourth) Ring}
\label{sec_fourth_ring}
Our best-fit model to the light-polarizing dust offers a good match to the $\lambda \gtrsim 50\mu$m SED (Figure \ref{sedmodel}).  However, there is remnant excess emission between 8--30 $\mu$m that is not reproduced by our dust model. The presence of warm circumstellar material well within 100 AU has been inferred not only from the above-mentioned SED fitting by \cite{Thi14} and \cite{Mawet17}, but also from CO observations \citep{Merin04, Goto06, Fisher00,Thi14,Flaherty16,White16}. Our detection of an inner clearing within the 44~AU ring, and the lack of polarization signal from sub-micron-sized grains (Section \ref{sec_amin}), imposes new constraints on the spatial extent of the warm dust responsible for the excess thermal emission at shorter wavelengths.

To account for this missing flux, we employ a simple model in MCFOST assuming the same grain size distribution, $a_{\text{min}}=4\mu$m, $a_{\text{max}}=1$ mm, porosity of 0\% and magnesium-rich olivines from our best-fit model. Motivated by \cite{Thi14}, the dust is characterized by a radial density distribution $R\propto r^{p}$ with $p=1$, no flaring ($\beta = 1$) and an inner disk radius of $R'_{in}=5$ AU. We keep the outer radius $R'_{out}$, and dust mass $M'_{dust}$ of the innermost disk as free parameters.

The best-fit thermal model for the innermost disk (light blue dashed line in Figure \ref{sedmodel}) indicates 300 K dust with a mass of $10^{-8}M_{\odot}$ ($3\times10^{-3}M_\oplus$) ranging from $R'_{in}=5.0$ AU up to $R'_{out}=15$ AU. This is well within the coronagraph IWA of GPI, and so presently undetectable in scattered light.

\section{Conclusion}\label{conclusions}
We have presented the first polarimetric detection of the inner 44 AU disk component of the pre-main sequence star HD 141569A. $H$-band polarimetric differential imaging with GPI has revealed a non-uniform ring-shaped optically-thin dusty disk inwards to 0$\farcs$25, at the highest signal-to-noise ratio attained to date. We find that the disk can be described radially with a combination of two power laws that peaks at $44$ AU and extends out to $100$ AU.
The disk also features an arc-like overdensity along the southern part that is reminiscent of the spiral arm structures previously known at larger scales in this system. The existence of this inner spiral arm structure and its co-location with CO emission detected by ALMA indicates that this may be a site of on-going icy grain destruction, perhaps driven by an unseen planetary perturber. The best-fit model to our polarimetry data indicates an optically-thin disk with a maximum surface density of $r_{max}\simeq R_{c}=44_{-12}^{+8}$ AU, a steep inner gradient ($\alpha_{\text{in}}=14$), and a shallower outer exponent ($\alpha_{\rm out}=-1.0_{-1.0}^{+0.5}$). The polarimetric observations are best described by a dust population model with a minimum size of $4_{-2}^{+4}\mu$m and a mass of $(1.0\pm 0.4)\times 10^{-6}M_{\odot}$ for non-porous grains up to 1~mm in size. A significant population of sub-micron grains is independently excluded by the lack of negative signal in the $H$-band Stokes $Q_{r}$ image. We use the thermal emission from our best-fit Mie model to estimate the amount of unseen dust inwards of 28~AU.  We find that a fourth innermost dust population, potentially a 5--15~AU belt, is required to fully reproduce the 8--30\micron\ SED under Mie theory assumptions.

With our new high-SNR polarimetric detection of the 44~AU ring, the richness of the circumstellar environment around HD 141569A can be appreciated under a new light. Considering resolved imaging data from other high-contrast facilities, the HD 1415169A debris disk shapes up to be made of at least three, and potentially four nested rings, with spiral structures on the three spatially resolved rings.  As such, it is an excellent laboratory for studying dynamically perturbed disks.
\acknowledgments
\textbf{Acknowledgments:} 
We thank our referee for a very insightful feedback on the dust considerations in our modeling.
This research was supported in part by a Discovery Grant by the Canadian Natural Sciences and Engineering Council (NSERC) to S.M., and by NSF grant AST-1413718 (GD). P.K. and J.R.G. thank support from NSF AST-1518332, NASA NNX15AC89G and NNX15AD95G/NEXSS. This work benefited from NASA's Nexus for Exoplanet System Science
(NExSS) research coordination network sponsored by NASA's Science Mission Directorate. Portions of this work were performed under the auspices of the U.S. Department of Energy by Lawrence Livermore National Laboratory under Contract DE-AC52-07NA27344.

\facilities{Gemini:South(GPI)}
\bibliography{HD141569bib}

 \end{document}